\documentclass[twocolumn,aps,prr,superscriptaddress,10pt,longbiblography]{revtex4-2} 

\usepackage{dcolumn}
\usepackage{amsmath}
\usepackage{multirow}
\usepackage{mathtools}
\usepackage{bbold}
\usepackage{hyperref}
\hypersetup{colorlinks=true,allcolors=blue}
\usepackage{tikz}
\newcommand{\bra}[1]{\langle {#1} |}
\newcommand{\ket}[1]{| {#1} \rangle}
\newcommand{\refEq}[1]{Eq.~(\ref{#1})}
\newcommand{\refFig}[1]{Fig.~\ref{#1}}

\newcommand{\refTab}[1]{Tab.~\ref{#1}}
\newcommand{\citeRef}[1]{Ref.~[\onlinecite{#1}]}
\DeclareMathOperator{\Tr}{Tr}
\newcommand{\expval}[1]{\langle {#1} \rangle}
\begin{document}

\title{Generalized Josephson effect with arbitrary periodicity in quantum magnets} 

\author{Anshuman Tripathi}
\affiliation{I.~Institut f{\"u}r Theoretische Physik, Universit{\"a}t Hamburg, Hamburg, Germany}

\author{Felix Gerken}
\affiliation{I.~Institut f{\"u}r Theoretische Physik, Universit{\"a}t Hamburg, Hamburg, Germany}
\affiliation{The Hamburg Centre for Ultrafast Imaging, Hamburg, Germany}

\author{Peter Schmitteckert}
\affiliation{HQS Quantum Simulations GmbH, Karlsruhe, Germany}

\author{Michael Thorwart}
\affiliation{I.~Institut f{\"u}r Theoretische Physik, Universit{\"a}t Hamburg, Hamburg, Germany}
\affiliation{The Hamburg Centre for Ultrafast Imaging, Hamburg, Germany}

\author{Mircea Trif}
\affiliation{International Research Centre MagTop, Institute of Physics, Polish Academy of Sciences, Warsaw, Poland}

\author{Thore Posske}
\affiliation{I.~Institut f{\"u}r Theoretische Physik, Universit{\"a}t Hamburg, Hamburg, Germany}
\affiliation{The Hamburg Centre for Ultrafast Imaging, Hamburg, Germany}

\begin{abstract}
\noindent Easy-plane quantum magnets are strikingly similar to superconductors, allowing for spin supercurrent and an effective superconducting phase stemming from their $U(1)$ rotation symmetry around the $z$-axis. 
We uncover a generalized fractional Josephson effect with a periodicity that increases linearly with system size in one-dimensional spin-$1/2$ chains at selected anisotropies and phase-fixing boundary fields.
The effect combines arbitrary integer periodicities in a single system, exceeding the $4\pi $ and $8\pi $ periodicity of superconducting Josephson effects of Majorana zero modes and other exotic quasiparticles. 
We reveal a universal energy-phase relation and connect the effect to the recently discovered phantom helices.
\end{abstract}

\maketitle

\section{Introduction}
\noindent The conventional Josephson effect describes a $2\pi$ periodic supercurrent flowing between two weakly coupled superconducting leads, whose magnitude is determined by the phase difference between the leads' condensates\cite{josephson1962PossibleNewEffects, golubov2004CurrentphaseRelationJosephson}.
Low-energy states at the junction potentially change this property by cycling through admissible excitations during an adiabatic change \cite{Kato1950OnTheAdiabaticTheoremOfQuantumMechanics, Berry1984, Laughlin1981}. 
Despite special trivial systems with this feature exist \cite{chiu2019FractionalJosephsonEffecta}, higher-periodic `fractional' Josephson effects have been theoretically and experimentally chased as signatures of topological quasiparticles. This includes the $4\pi$ Josephson effect of Majorana modes \cite{Kwon2004, LutchynSauDasSarma2010, kurter2015EvidenceAnomalousCurrent, wiedenmann20164pperiodicJosephsonSupercurrent}, where the flux is altered by two quanta instead of one quantum in order to recover the initial state of the system, and the $8\pi$ Josephson effect or more intricate behavior for interacting systems and general non-abelian anyons \cite{Jiang2011UnconventionalJosephsonSignaturesMajoranaBoundStates, PedderMengTiwariSchmidt2017MissingSharpiraStepsAndThe8PiPeriodicJosephsonEffect, NayakSternFreedmanDasSarma2008NonAbelianAnyonsAndTopologicalQuantumComputation, Trif2021PhysRevResearch.3.013003}. 
Additionally, fractional Josephson-like effects, pertaining to the transport of spin rather than charge, have been predicted in several electronic setups that harbor Majorana modes \cite{JiangPRB2013, MengPRB2013, KotetesJKPS2013}. 
Focusing on the main aspects of a $U(1)$ phase symmetry and supercurrent transport, a generalized Josephson setup uses a general one-dimensional quantum medium with fixed phases at its terminals. Easy-plane quantum magnets with a $U(1)$ rotation symmetry around the $z$-axis accommodate those features \cite{Nogueira_2004, tserkovnyak2020QuantumHydrodynamicsSpin, gerken2023AllProductEigenstates, Hoffman2023SpuerfluidTransportInQuanutmSpinChains} by their in-plane superfluid phase $\phi$ and the associated spin supercurrent $C^z_{j,k}$ between sites $j$ and $k$ that changes the local magnetization $\partial_t S^z_j = \sum_{k} C^z_{j,k}$.

An aperiodicity upon periodic drive in general classical physical systems is not surprising. For instance, a classical ribbon fixed at one end accumulates twists when rotated at the other end. 
Motivated by that, we consider an easy-plane $XXZ$ one-dimensional ferromagnet controlled by magnetic fields at the boundaries where the superfluid transport of spin current resembles the one of electric supercurrent in superconductors \cite{WindingThorePosske2019, KimTserkovnyak2016TopologicalEffectsOnQuantumPhaseSlipsInSuperfluidSpinTransport, tserkovnyak2020QuantumHydrodynamicsSpin}. 
There, winding up a quantum spin helix is facilitated for special values of the axial anisotropy \cite{WindingThorePosske2019} that are connected to roots of unity in the case of spin-1/2 sites \cite{Nepomechie2003b, pasquier_common_1990}. 
Furthermore, exact solutions with the chiral Bethe ansatz have put easy-plane boundary-controlled spin chains in the focus, prominently because of phantom helices \cite{Nepomechie2003, Nepomechie2004Addendum, Cao2003, Cao2013, PopkovPhantomBetheRoots, PopkovChiralCoordinateBetheAnsatz2021PhysRevB.104.195409}, which are high-energy helical spin product states that have recently been realized in cold atoms \cite{jepsenLonglivedPhantomHelix2022}. These helices possess increased stability against environmental influence \cite{jepsenLonglivedPhantomHelix2022, Popkov2013PhysRevE, Popkov2017PhysRevA}, and could experience protection when employed as qubits \cite{Popkov2024ChiralBasisForQbits}.
Phantom helices are accompanied by helical states that maximize the spin current at spectral degenerate points, that equally experience helical protection against local magnetic perturbations \cite{Kuehn2023}. 

In this work, we show that the generalized Josephson effect when adiabatically winding up spin helices in easy-plane spin-$1/2$ chains has an integer periodicity for special values of the Heisenberg anisotropy $\Delta$ that increases linearly with the length of the system.  
The winding-up, which adds one helical quasiparticle excitation by each rotation in real space, is most pronounced when $\Delta$ is half the ferromagnetic exchange coupling $J$.   
We furthermore show that the spin-current-maximizing helices that are created during the winding successively become degenerate to all phantom helices, thereby connecting the two types of helices. Our numerical results for chain lengths with up to $20$ quantum spins asymptotically collapse onto a universal curve, the universal energy-phase relation of the $Z_n$ Josephson effect. 
We explain this by a universal effective $n$-state hopping model in angular momentum space.

The structure of this article is as follows. In Sec.~\ref{Sec:2}, we introduce the model for the spin-$1/2$ Heisenberg XXZ chain and the numerical method employed to adiabatically twist the boundary fields.~In Sec.~\ref{Sec:3}, we examine the wound-up helix at the specific anisotropy $\Delta = -|J|/2$, highlighting the linear relationship between the $Z_n$ effect and the system length. The cumulative effects of this adiabatic winding lead to a universal energy curve, which we elaborate in Sec.~\ref{Sec:4} through the lens of an effective hopping model. In Sec.~\ref{Sec:5}, we investigate the adiabatic cycles of additional states and finally conclude with an outlook in Sec.~\ref{Sec:6}.

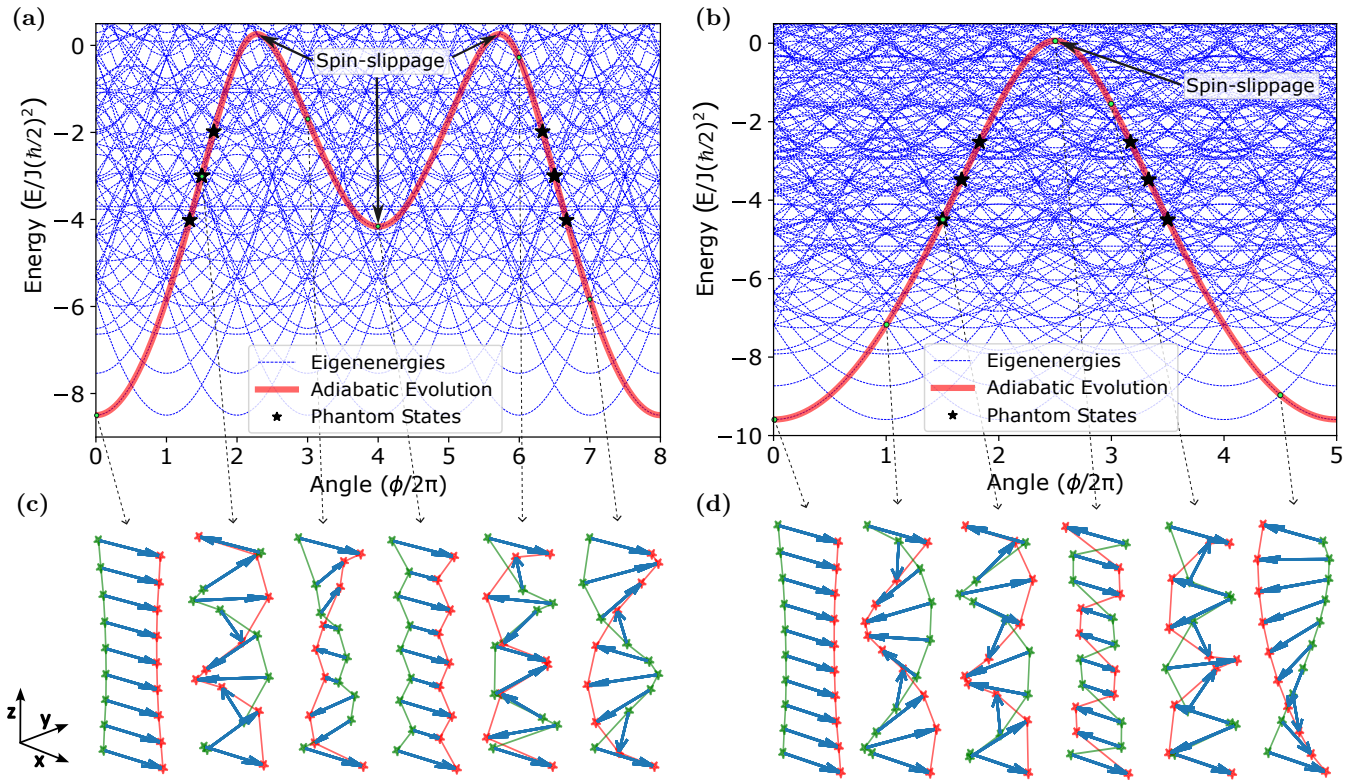
\begin{figure*}[t]
        \centering
        \setlength{\unitlength}{0.1\textwidth}
        \begin{picture}(10,5.9)
            \put(0,0){\includegraphics[width=0.99\textwidth]{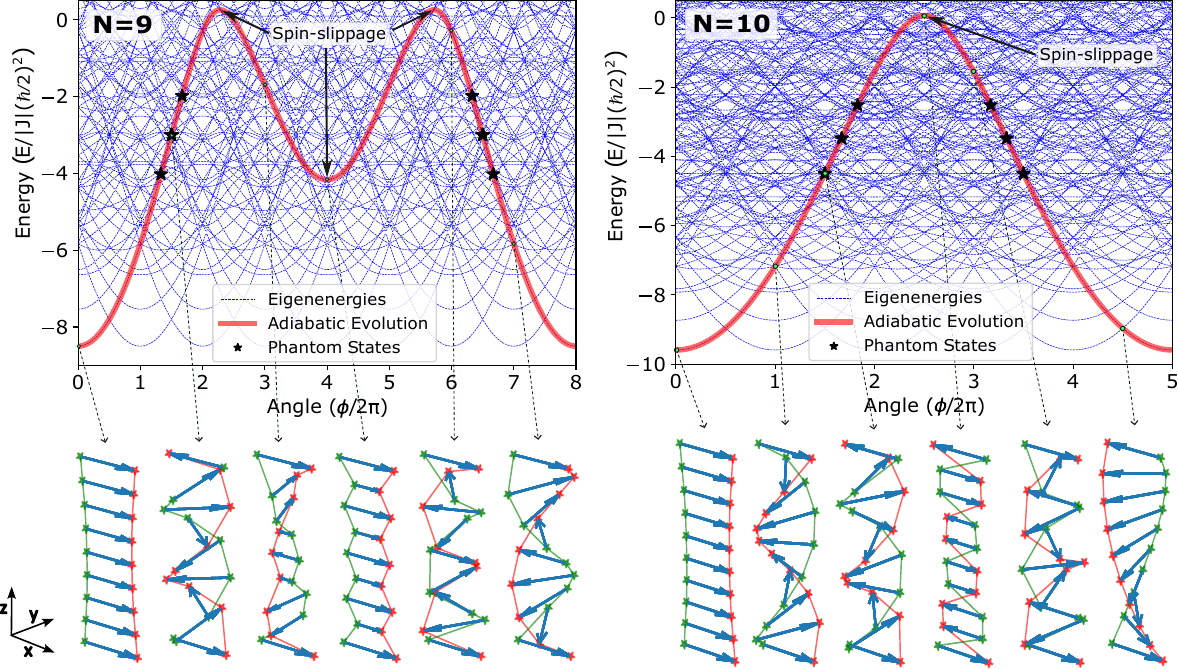}}
            \put(0.03,5.6){\textbf{(a)}}
            \put(5.1,5.6){\textbf{(b)}}
            \put(0.06,2){\textbf{(c)}}
            \put(5.1,2){\textbf{(d)}}
        \end{picture}        
    \caption{$Z_n$ Josephson effect in the $XXZ$ Heisenberg chain, i.e., the adiabatic evolution of the ground state (red curve) where the first spin is fixed while the last spin is rotated by a phase $\phi$. The instantaneous eigenenergies are dotted blue lines in the energy-phase diagram, and the black stars mark the phantom helical states. The adiabatic curve passes through all the degenerate eigenspaces of phantom states. Panel (a) exemplifies the $Z_8$ effect for chain length $N=9$ and (b) the $Z_5$ effect for chain length $N=10$. Panel (c) and (d) show spin expectation values at selected rotation phases.}
    \label{fig:AdiabaticCurve}
    \vspace{-1em}
\end{figure*}

\section{Model and methods} \label{Sec:2} 
\noindent We consider the one-dimensional $XXZ$ Heisenberg model consisting of $N$ quantum spins $\mathbf{S}_j$ with $S = \hbar/2$. 
We polarize the boundary spins into the directions of large boundary fields which locally breaks the $U(1)$ symmetry, e.g., by external magnetic fields, ferromagnetic islands, or leads. 
The effective low-energy Hamiltonian after projecting out the polarized boundary spins is (see Appendix \ref{app:LowEnergyHamil})
\begin{align} \label{eq:ChainHam} 
    H(\phi) = &\sum_{j=2}^{N-2}\left[J(S_j^{x} S_{j+1}^{x} +S_{j}^{y} S_{j+1}^{y}) + \Delta S_{j}^{z} S_{j+1}^{z}\right]  \nonumber \\
    &+ J\left[\mathbf{\tilde{S}}_1\cdot\mathbf{S}_2 +  \mathbf{S}_{N-1}\cdot\mathbf{\tilde{S}}_N(\phi)\right]. 
\end{align}
Here, $J$ is the exchange coupling, and $\Delta$ is the axial Heisenberg anisotropy. The $\mathbf{S}_j$ for $2\le j\le N-1$ are quantum spin-1/2 operators, $\mathbf{\tilde{S}}_1 = (1,0,0)\cdot\hbar/2$ and $\mathbf{\tilde{S}}_N(\phi) = (\cos{\phi}, \sin{\phi}, 0)\cdot\hbar/2$ are the directions of the terminal magnetic fields. 

To study the winding up of helices when $\phi$ is adiabatically changed, we implement the Hamiltonian in \refEq{eq:ChainHam} numerically and calculate its eigenstates with the Lanczos algorithm \cite{LanczosBrezinski1997ProjectionMF}. 
To compute the adiabatic evolution, we use two independent implementations with agreeing results. For small systems, we follow Kato's projection formalism \cite{Kato1950OnTheAdiabaticTheoremOfQuantumMechanics} with a small discrete step size $\delta \phi$ to find the adiabatically evolved states starting from the ground state (data of Fig.~\ref{fig:AdiabaticCurve}), see Appendix \ref{app:KatoApproach} for further details.
The second approach, tailored towards the simulation of larger systems, follows the adiabatic state tracking method \cite{AdiabaticTrackingPeterPhysRevLett.111.120602} combined with a projector onto an energy interval via a contour integration (FEAST) (data of Fig.~\ref{fig:UniversalScaling}) \cite{DMRGPolizziPhysRevB.79.115112}. 
Specifically, we project the Hamiltonian on an energy interval around the next energy based on a linear prediction of the previous energies. Then, we select the adiabatically evolved state via the overlap with the last state and perform a sparse matrix diagonalization taking the selected state as an initial guess to improve the eigenvector. 

\section{$Z_n$ effect} \label{Sec:3}
\noindent At roots of unity, i.e., values of the axial Heisenberg anisotropy $\Delta = -|J| \cos{(\pi/k)}$ with odd integer $k<N$, relevant energy gaps close, and the ground state can evolve to excited states in the adiabatic limit \cite{WindingThorePosske2019}. 
We find that the ordered ground state gradually twists into a helix with increasing winding number until the helix starts to unwind and returns to its original state due to the finite size of the quantum spin system, see Fig.~\ref{fig:AdiabaticCurve}. 
We note that for each Hamiltonian that is ferromagnetic in the $xy$ plane, we can construct an antiferromagnetic one through the unitary transformation $U = \prod_j^N e^{-(i/\hbar)\pi (j-1) S_j^z} $ which flips alternating spins, effectively changing the sign of $J$ and vice versa.
We refer to the number of twists it takes to return to the initial state as periodicity, and to a periodicity of $2\pi n$ as the generalized $Z_n$ Josephson effect. 
Here, we focus on $\Delta = -|J|/2$, for which we find the longest periods for the states among all Heisenberg anisotropies. 
For example, the ground states of the spin chains with $N=9$ and $N=10$ are $Z_8$ and $Z_5$ periodic, respectively, as shown in \refFig{fig:AdiabaticCurve}. 
At the start and the middle of the adiabatic curve, there are time-inversion symmetric points after which further winding up acts like unwinding, i.e., $|\psi (\phi) \rangle = U |\psi( - \phi)\rangle$ with $U = \prod_j^N e^{-(i/\hbar)\pi S_j^x}$. This corresponds to mirroring each spin about the $xz$ plane, which inverts the helicity of a state, see the Supplemental Video for demonstration. 
From chain lengths up to $N=20$, we find that the periodicity increases linearly with the chain length according to 
\begin{align}\label{eq:PeriodicityGS}
    P(N) = \begin{cases}
    N/2, &\text{ for even $N$,}\\
    N-1, &\text{ for odd  $N$.}
    \end{cases}
\end{align}
The periodicity is twice as large as the preceding chain length for odd lengths, which originates from an extra Kramer's degeneracy in odd chains at half phase-rotation \cite{KleinKramer1952, Kramer1930, Trif2021PhysRevResearch.3.013003}.
The wound-up quantum spin helices before spin slippage are spin current maximizing helices at their degenerate eigenspaces, carrying the maximum amount of spin current along the direction of the spin chain $C_{j,j+1}^z = (\mathbf{S}_j \times \mathbf{S}_{j \pm 1})^z$ \cite{WindingThorePosske2019, Kuehn2023}. 
For each eigenstate, the spin current is spatially homogeneous as $\partial_t \langle S_j^z \rangle = 0$, which implies $\langle C_{j-1,j}^z \rangle = \langle C_{j,j+1}^z \rangle$ for all sites $j$, reflecting the spin current's continuity equation.
Interestingly, the adiabatic winding neither results in a net $z$-spin transfer nor spin current between the edges of the chain. The process therefore is distinct from a Thouless quasiparticle pump \cite{Thouless1983PhysRevB.27.6083, citro_thouless_2023} or its extensions in adiabatic spin transfer \cite{FuKane2006PhysRevB.74.195312} albeit carrying similarities. To see this, note that the $z$-spin expectation values for each spin vanish as the boundaries are polarized in the $xy$ plane and $|\Delta|<|J|$, i.e., $\langle S_j^z \rangle = 0$ at all times.
In addition, the adiabatic contribution to the transferred spin current vanishes due to its special form 
\begin{align}
\label{eq:SSC}
\langle C_{j,j+1}^z\rangle =  
\langle \left(\mathbf{S}_j \times \mathbf{S}_{j + 1}\right)^z \rangle = -
\langle \partial_\phi H\rangle,
\end{align}
where we have combined the above continuity equation with the rotating boundary term, see Appendix \ref{app:SpinPump} for details. 
By \refEq{eq:SSC}, the current-phase relation can readily be obtained by differentiating the energy-phase relation in \refFig{fig:AdiabaticCurve}.

Unexpectedly, we find a relation of the $Z_n$ periodic Josephson effect to the recently introduced phantom helical states \cite{PopkovPhantomBetheRoots, PopkovChiralCoordinateBetheAnsatz2021PhysRevB.104.195409}.
Phantom states are the product eigenstates of the Hamiltonian in Eq.~(\ref{eq:ChainHam}) when the phantom boundary conditions 
\begin{align}
    \phi_{ph} = \pm (N-M)\pi/3 + \pi \delta_{3,M}  \text{ mod } 2\pi
\end{align}
are satisfied, where  $M$ can be chosen as $1, 3, \text{ or } 5$ \cite{Kuehn2023}. 
These states have the energy $E_{ph} = \left(N-M\right) \Delta \frac{\hbar^2}{4}$. 
We observe that the adiabatically evolved state reaches the degenerate subspaces of the phantom states for all chain lengths, see \refFig{fig:AdiabaticCurve}, but not the phantom states themselves, see Appendix \ref{app:Overlaps} for a list of overlaps. 

\section{Asymptotic universal scaling} \label{Sec:4}
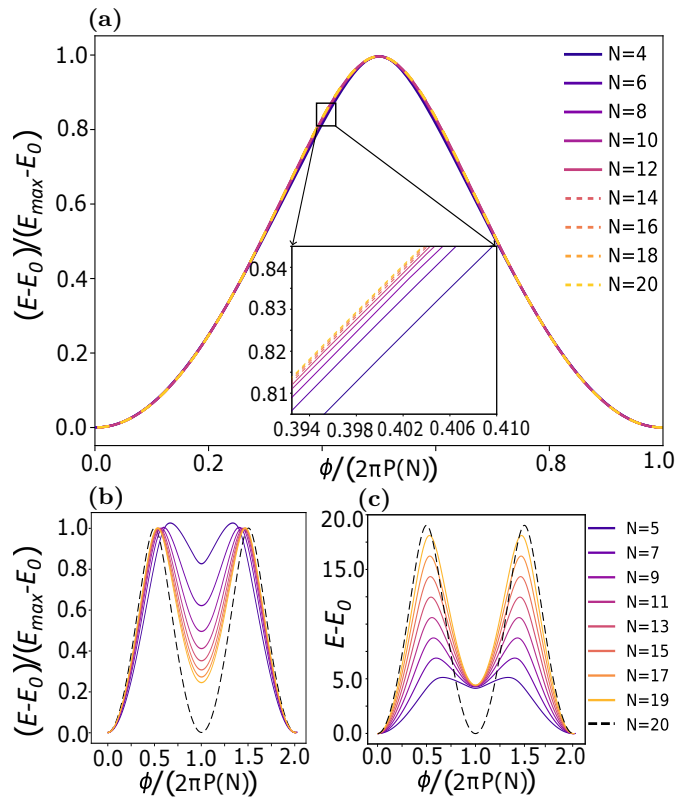
\begin{figure}[t]
    \vspace{-2em}
    \centering
        \setlength{\unitlength}{0.1\textwidth}
        \begin{picture}(5.2,6.2)(0,0)
            \put(0,0){\includegraphics[width=0.49\textwidth]{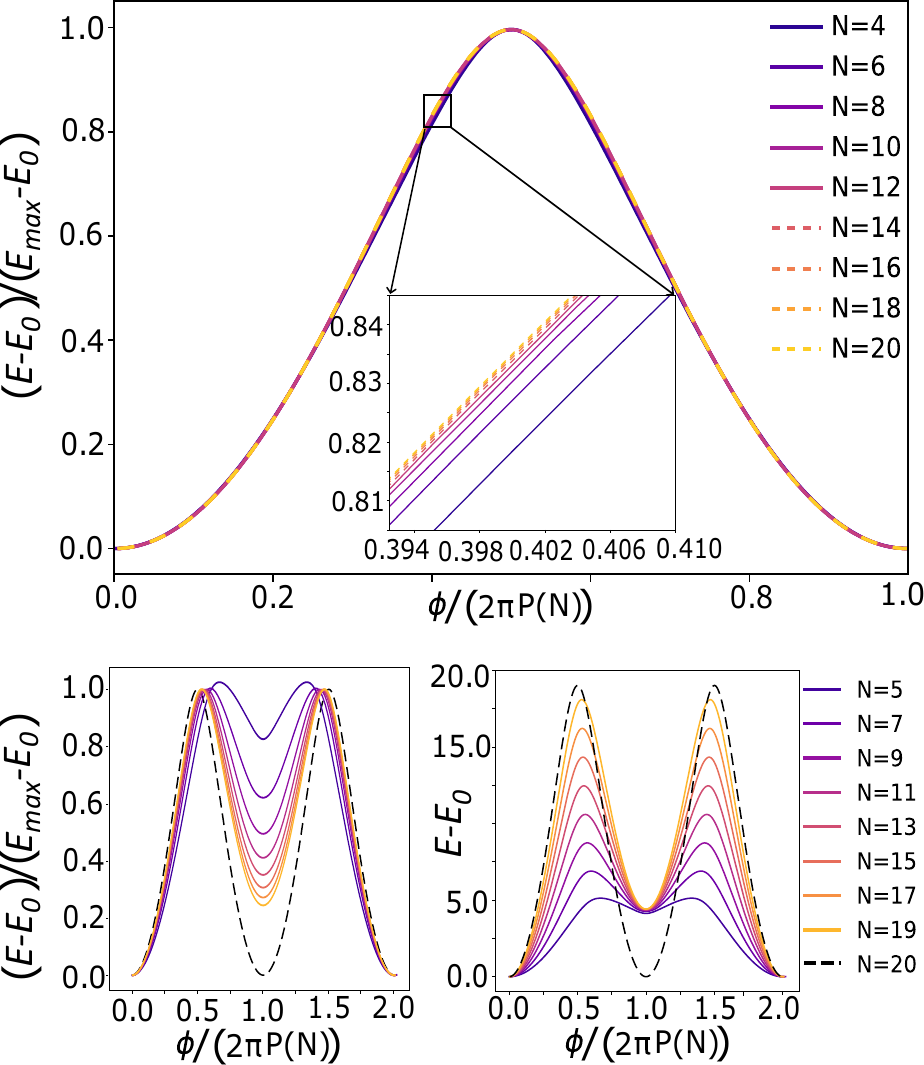}}
            \put(0.55,5.72){\textbf{(a)}}
            \put(0.55,2.17){\textbf{(b)}}
            \put(2.55,2.16){\textbf{(c)}}
        \end{picture}
        
    \caption{Asymptotic universal energy-phase relation of the $Z_n$ Josephson effect at $\Delta = -|J|/2$ for chain lengths up to $N=20$. (a) Asymptotic universal curve for even chains after scaling the phase by the system's periodicity in \refEq{eq:PeriodicityGS} and the energy by the maximally pumped energy. (b) Scaled curve including odd chain lengths. (c) Convergence of the time-reversal symmetric point's energy for odd chains when only rescaling the phase.}
    \label{fig:UniversalScaling}
\end{figure}

\noindent The phantom energies' linear increase in the chain length and the linear increase in the ground state periodicity point towards a general scaling behavior of the energy-phase relation. \refFig{fig:UniversalScaling}(a) shows the collapse for even chain lengths onto this universal curve. To this end, the phase $\phi$ is scaled according to the periodicity in Eq.~(\ref{eq:PeriodicityGS}), and the energy difference to the ground state $E-E_0$ is scaled by its maximum during the adiabatic evolution, which follows $E_{max}-E_0 \approx 0.935N+0.335$ ($\sigma = 0.023$), see Appendix \ref{app:Fits}. 
The inset in \refFig{fig:UniversalScaling}(a) shows the convergence for long chains, where small chains deviate more from the asymptotic curve, likely caused by reduced entanglement at the boundary. 
After rescaling correspondingly, odd chains are indicated to converge to the same universal curve in the thermodynamic limit, see \refFig{fig:UniversalScaling}(b), but from their absolute values, they converge to a finite universal energy at the time-reversal symmetric point, see \refFig{fig:UniversalScaling}(c), and follow the $1/N$ scaling of this finite value, see Appendix \ref{app:Fits}. 
\begin{figure}
    \centering
    \includegraphics[width=1\linewidth]{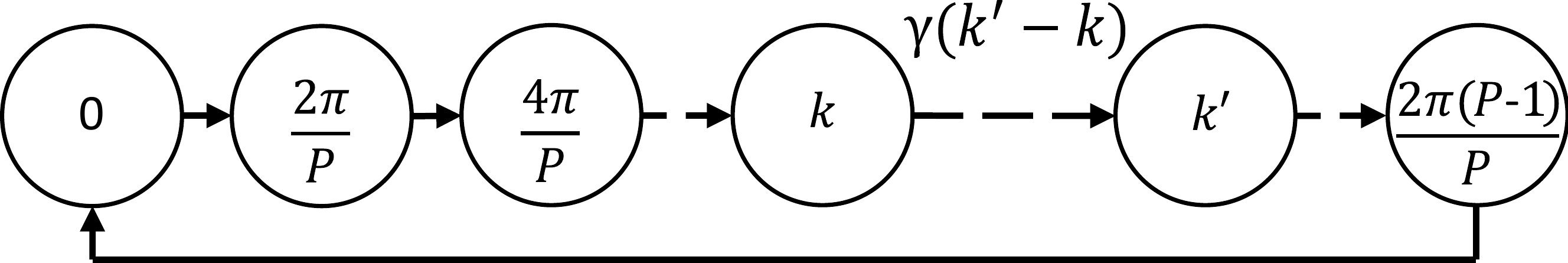}
    \caption{The helical states after an integer number of rotations are the eigenstates of a translationally invariant $P$-site hopping model in angular momentum space in \refEq{eq:RotarModel}, whose hoppings follow the Fourier transform $\gamma$ of the universal energy-phase relation.}
    \label{fig:PeriodicHopping}
    \vspace{-1em}
\end{figure}

The universal scaling enables us to develop a universal hopping model in the corresponding angular momentum Fourier space of $\phi$. To this end, we rewrite the Hamiltonian in Eq.~(\ref{eq:ChainHam}) in diagonalized form at $\phi = 0$, 
\begin{align} \label{eq:DiagonalisedHamiltonian}
    \frac{H-E_0}{E_{\text{max}}-E_0} = \sum_{h \in Q} \alpha_{h} | h \rangle \langle h | + \sum_{\varepsilon \neq h} \alpha_{\varepsilon} | \varepsilon \rangle \langle \varepsilon |, 
\end{align}
where $\alpha_{h}$ and $\alpha_{\varepsilon}$ denote the rescaled energy of the eigenstate $|h \rangle$ and $|\varepsilon \rangle$, and $Q=\{ 0,1,2, \cdots, P-1\}$ labels the helical states after a $2\pi h$ rotation. By introducing the Fourier transform $|k \rangle = \sum_{h \in Q} e^{-ikh}|h \rangle /\sqrt{P}$ with $k \in (2\pi/P) Q$, we arrive at the translationally invariant $P$-site hopping model
\begin{align} 
     \frac{H-E_0}{E_{\text{max}}-E_0} =& \sum_{k,k'} \gamma(k' - k)|k' \rangle \langle k | + R, \label{eq:RotarModel}
\end{align}
where the remainder $R$ includes all nonuniversal states and $\gamma(k'-k) = \sum_{h\in Q} e^{-ih (k'-k)}\alpha_h/P$ is the Fourier transform of the  universal energy-phase relation of \refFig{fig:UniversalScaling}(a) that describes the universal hoppings, see \refFig{fig:PeriodicHopping}.

\section{Super-winding states} \label{Sec:5}
\begin{figure}[t]
    \centering
    \setlength{\unitlength}{0.1\textwidth}
        \begin{picture}(5,7.3)(0,0)
            \put(0,0){\includegraphics[width=0.49\textwidth]{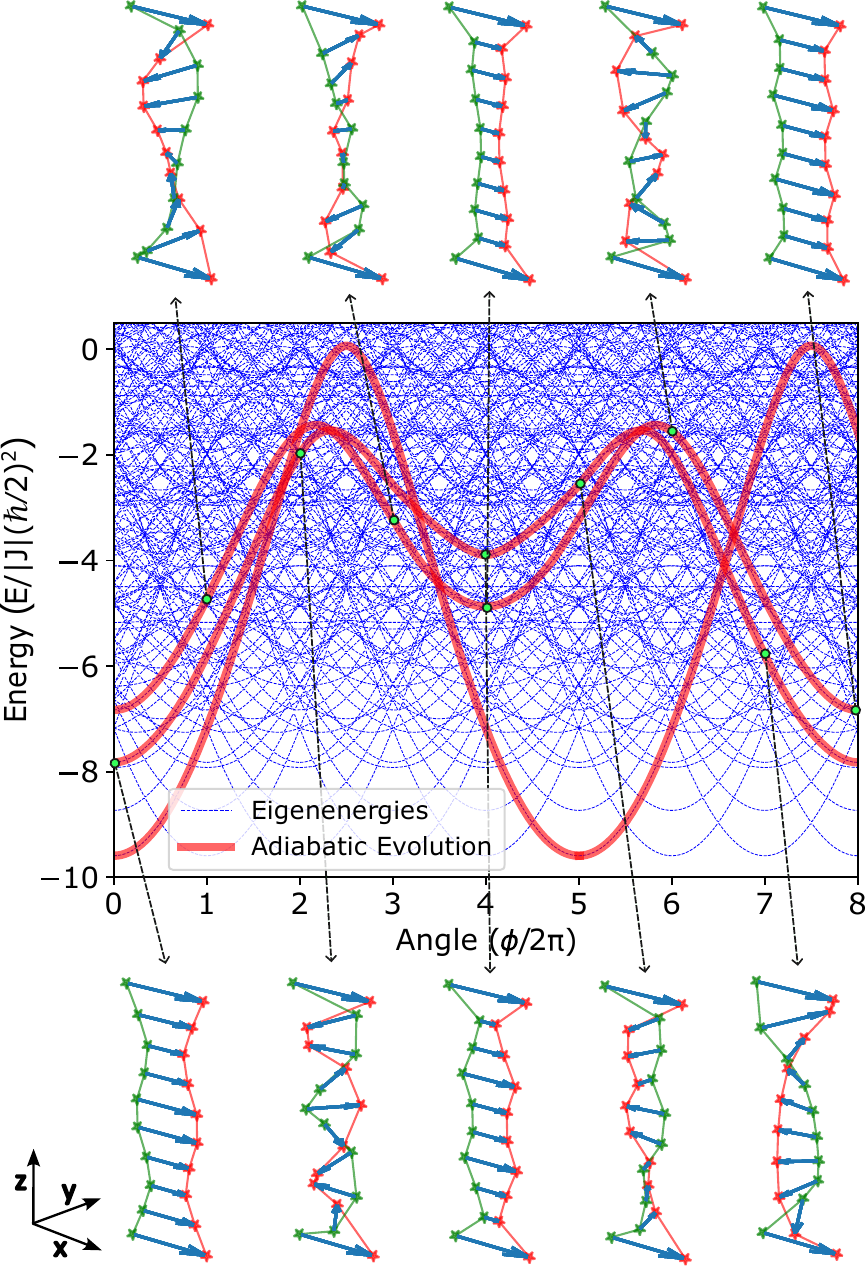}}
            \put(0.05,7.1){\textbf{(a)}}
            \put(0.05,5.3){\textbf{(b)}}
            \put(0.05,1.6){\textbf{(c)}}
        \end{picture}
        
    \caption{Super-winding Josephson effect of selected excited states (red) for chain length $N=10$. (b) $Z_8$ periodicity for the third and eighth excited states, exceeding ground state periodicity. (a) and (c) spin expectation values at selected $\phi$.}
    \label{fig:SuperWinding}
    \vspace{-2em}
\end{figure}
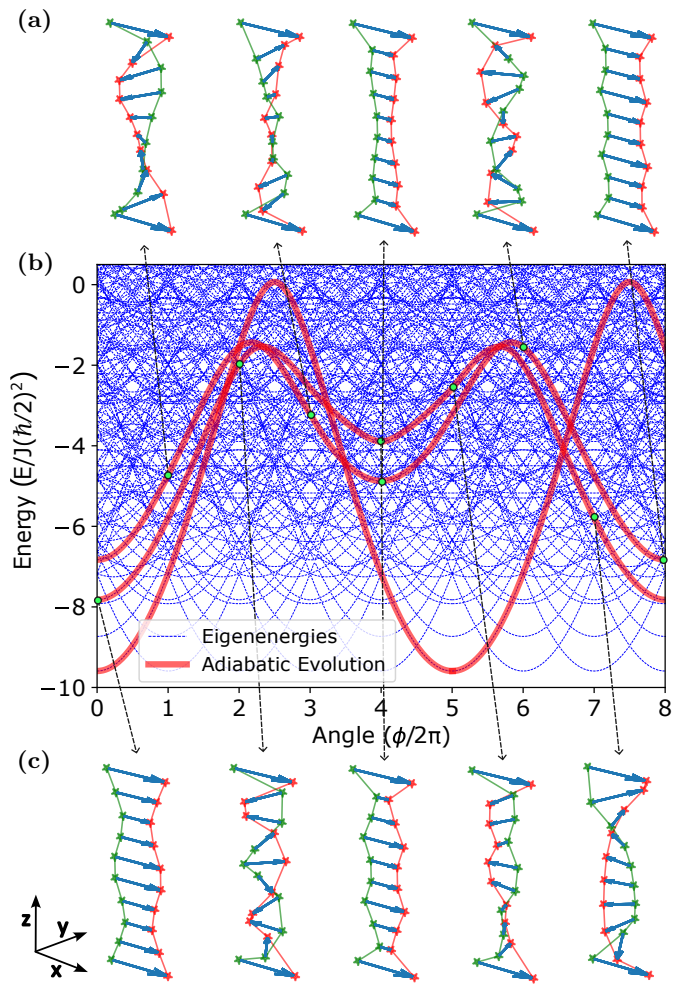

\noindent Peculiarly, the $Z_n$ Josephson effect of the ground state is generally not the one with the largest periodicity of the many-body spectrum. To find these super-winding states, we calculate the permutation matrix that maps the instantaneous eigenstates at $\phi = 0 $ to those reached after an adiabatic evolution to $\phi =2\pi$ and analyze its cycles.  We observe that for chains with more than $7$ spins, there are excited states with a periodicity larger than that of the ground state. For example, \refFig{fig:SuperWinding} shows the super-winding states for chain length $N=10$. There, the third and eighth excited eigenstates are $Z_8$-periodic, whereas the ground state is $Z_5$-periodic. Furthermore, there are additional states with the same periodicity as the ground state, yet, the energy pumped into the system remains maximal for the ground state's Josephson effect. For longer chains, the number of super-winding states further increases, see Appendix \ref{app:PermutaionCycles} for a complete analysis of selected chain lengths.

\section{Conclusion and outlook} \label{Sec:6}
\noindent We describe a generalized fractional Josephson effect with arbitrary integer periodicity and a universal energy-phase relation in $XXZ$ Heisenberg chains that are adiabatically controlled by magnetic fields at their ends.
This extends the class of higher-periodic Josephson-like effects, where the system's boundaries break the $U(1)$ symmetry of a general one-dimensional quantum system \cite{Nogueira_2004} and generalizes the $4 \pi$ and $8 \pi$ charge and spin Josephson effects originating from Majorana zero modes \cite{MengPRB2013, JiangPRB2013,KotetesJKPS2013}.
Given the recent experimental breakthrough in realizing $XXZ$ Heisenberg systems with precisely tunable Heisenberg anisotropy in cold atom systems \cite{jepsenSpinTransportTunable2020}, our findings have direct experimental implications. Specifically, the two-component boson system with the lowest two hyperfine states of ${}^{\text{7}}\text{Li}$ in the Mott-insulating regime allows for the direct resolution of the time-dependent spin expectation values by the ensemble-averaged contrast \cite{jepsenSpinTransportTunable2020, jepsen2021TransverseSpinDynamics, jepsenLonglivedPhantomHelix2022, dimitrova2023ManybodySpinRotation}, connecting to our Figs.~\ref{fig:AdiabaticCurve} and \ref{fig:SuperWinding}. The crucial challenge here is to tilt the adiabatically controlled boundary fields in \citeRef{dimitrova2023ManybodySpinRotation} into the $xy$ plane.
Additionally, Rydberg atoms \cite{glaetzle2014QuantumSpinIceDimer, semeghini2021ProbingTopologicalSpin, kim2024RealizationExtremelyAnisotropic} and solid-state systems, where the universal behavior of the spin current is potentially directly accessible, can help to realize the system and next nearest neighbor coupling or tunable Dzyaloshinskii–Moriya interaction, e.g., by an external electric field \cite{Stepanov2017PhysRevLett.118.157201}, can achieve tunability to the sweet spots of the Heisenberg anisotropy. 
Furthermore, designed counter-diabatic winding protocols can stabilize the adiabatic evolution, e.g., time-dependent rotations of the magnetic boundary that suppress excitations away from adiabaticity \cite{CampoShortcutToAdiabacityPhysRevLett.111.100502, Aslani2023SuperpositionStatesViaShortcutToAdiabcity}.

In superconducting $Z_n$ Josephson effects \cite{Kwon2004, LutchynSauDasSarma2010, Jiang2011UnconventionalJosephsonSignaturesMajoranaBoundStates, PedderMengTiwariSchmidt2017MissingSharpiraStepsAndThe8PiPeriodicJosephsonEffect, NayakSternFreedmanDasSarma2008NonAbelianAnyonsAndTopologicalQuantumComputation, Trif2021PhysRevResearch.3.013003}, the adiabatic phase change is associated with the excitations of an exotic quasiparticle. Here, those excitations are well described by the universal hopping model in \refEq{eq:RotarModel} and we can calculate that the adiabatically accumulated $z$-spin and the $z$-spin current both vanish, see Appendix \ref{app:SpinPump}. This distinguishes our setup from the Thouless quasiparticle pump \cite{Thouless1983PhysRevB.27.6083} and its $\mathbb{Z}_2$ generalizations \cite{citro_thouless_2023, Shindou2003QuantumSP, Shiozaki2022PhysRevB.106.125108, FuKane2006PhysRevB.74.195312}. 
Yet, the helical quasiparticles share similarities to $Z_n$ symmetric parafermions in clock models \cite{SchmidtBosonizationParafermion2020, Alicea2018PhysRevB.98.085143, PedderMengTiwariSchmidt2017MissingSharpiraStepsAndThe8PiPeriodicJosephsonEffect} and Majorana modes \cite{MengPRB2013, JiangPRB2013,KotetesJKPS2013}. 
Like in Majorana modes, we have a $Z_2$ periodicity for small systems where rotating once creates a quasiparticle, and is annihilated by the second rotation, demonstrating that the wound-up excitation is its own antiparticle.
Ultimately, the energy in the $Z_n$ Josephson effect reaches the middle of the spectrum, which could be used in spin-based quantum battery applications \cite{vedmedenko2014TopologicallyProtectedMagnetic, le2018SpinchainModelManybody}.
There, we expect thermal fluctuations to have a detrimental impact on the feasibility of adiabatic winding for long chains which we propose as a subject for further studies.
We hope that our work stimulates further research to specify the nature of the corresponding quasiparticles and eventually detect them in quantum spin systems or their counterparts in the fractional Josephson effects of interacting electrons.  
 
\begin{acknowledgments} 
\noindent We thank V.~Popkov and T.~Schmidt for helpful discussions. A.T.~and T.P.~acknowledges funding by the European Union (ERC, QUANTWIST, Project number 101039098). The views and opinions expressed are however those of the authors only and do not necessarily reflect those of the European Union or the European Research Council, Executive Agency. F.G.,~M.Th.~and T.P.~acknowledge funding by the Cluster of Excellence ``CUI: Advanced Imaging of Matter'' of the Deutsche Forschungsgemeinschaft (DFG) – EXC 2056 – Project ID 390715994. T.P.~acknowledges funding of the DFG (German Research Foundation) Project No. 420120155. M.Tr.~acknowledges support from the National Science Centre (Poland) OPUS Grant No. 2021/42/B/ST3/04475, and the Foundation for Polish Science project ``MagTop'' (No. FENG.02.01-IP.05-0028/23) cofinanced by the European Union from the funds of Priority 2 of the European Funds for a Smart Economy Program 2021-2027 (FENG).
\end{acknowledgments}

\onecolumngrid 

\appendix

\section*{Appendix}

\noindent In this Appendix, we provide details of the numerical method used to track the adiabatic evolution, give details on the scaling analysis, list the overlap values of the phantom helices with the wound-up helix, and discuss the complete state cycles for selected spin chain lengths that we use to determine the super-winding state. Additionally, we derive the low-energy Hamiltonian for the $XXZ$ Heisenberg chain coupled with large boundary magnetic fields and derive the vanishing adiabatic contribution to the $z$-spin current. 

\section{Low-energy Hamiltonian} \label{app:LowEnergyHamil}

\noindent In this section, we derive the effective Hamiltonian that represents the reduced low-energy Heisenberg model of the main text. The full Hamiltonian of an XXZ Heisenberg spin chain with N spin-$1/2$ and boundary spins coupled with boundary magnetic fields $\mathbf{B}_1 = B \mathbf{n}_1 = B(-1,0,0) $ and $\mathbf{B}_N(\phi) =B \mathbf{n}_N = B(-\cos\phi, -\sin\phi,0) $ is 
\begin{align}
    H(\phi) = &\underbrace{ \sum_{j=1}^{N-1}\left[J(S_j^{x} S_{j+1}^{x} +S_{j}^{y} S_{j+1}^{y}) + \Delta S_{j}^{z} S_{j+1}^{z}\right]}_{\displaystyle H_{ex}} + \mu \left[\mathbf{B}_1\cdot\mathbf{S}_1  +  \mathbf{S}_{N}\cdot \mathbf{B}_N(\phi)\right] + \mu \hbar B \cdot \mathbb{1},
\end{align}
where $\mu$ is the coupling constant between the edge spins and the magnetic fields and $H_{ex}$ refers to the bulk part of the Hamiltonian consisting only of exchange interactions. For regularization purposes, we added a constant term $ \mu \hbar B $ in the Hamiltonian. For $B \rightarrow \infty$ we can write the eigenvalue equation as 
\begin{align}
    \lim_{B \rightarrow \infty} \hspace{1em} H_{ex} + \mu B \left[ \mathbf{n}_1\cdot\mathbf{S}_1 + \mathbf{S}_N\cdot \mathbf{n}_N(\phi) + \hbar \mathbb{1} \right]| \psi \rangle &= E_\psi |\psi \rangle  \nonumber \\
    \implies  \mu  \left[ \mathbf{n}_1\cdot\mathbf{S}_1 + \mathbf{S}_N\cdot\mathbf{n}_N(\phi) + \hbar \mathbb{1} \right] | \psi \rangle &\simeq  \varepsilon_\psi | \psi \rangle, \hspace{2em} \text{with } \varepsilon_\psi = \frac{E_\psi}{B}. 
\end{align} 
We hence only get finite energies in the Hamiltonian $H(\phi)$ when $\varepsilon_\psi = 0$ and the edge spins are perfectly polarized as $|-\rangle_1 = \frac{1}{\sqrt{2}} (| \uparrow \rangle + | \downarrow \rangle)$ and $|-\rangle_N = \frac{1}{\sqrt{2}} (| \uparrow \rangle + e^{i \phi} | \downarrow \rangle)$. For other states, the energy $E_\psi$ diverges for $B \rightarrow \infty$. We therefore project to this low-energy subspace using the projection operator   
\begin{align}
    P  = \prescript{}{1}{|-\rangle} \langle-|_1 \otimes \mathbb{1}_2 \otimes \cdots \mathbb{1}_{N-1} \otimes \prescript{}{N}{|-\rangle} \langle -|_N.
\end{align}
The projected Hamiltonian in this subspace is 
\begin{align} 
    P^\dagger H(\phi) P =& - \frac{\mu \hbar B}{2} P - \frac{\mu \hbar B}{2}  P + \mu \hbar B P + JP^\dagger( S_1^x S_2^x + S_1^y S_2^y +S_{N-1}^x S_N^x + S_{N-1}^y S_N^y )P \nonumber \\ 
    & +\Delta P^\dagger( S_1^z S_2^z + S_{N-1}^z S_N^z )P^\dagger + \sum_{j=2}^{N-2} P^\dagger \left[J(S_j^{x} S_{j+1}^{x} +S_{j}^{y} S_{j+1}^{y}) + \Delta S_{j}^{z} S_{j+1}^{z}\right] P, \nonumber  \\
     =& \hspace{0.5em} J \frac{\hbar}{2} (S_2^x + \cos{\phi} S_{N-1}^x + \sin{\phi} S_{N-1}^y) P +\sum_{j=2}^{N-2} \left[J(S_j^{x} S_{j+1}^{x} +S_{j}^{y} S_{j+1}^{y}) + \Delta S_{j}^{z} S_{j+1}^{z}\right] P.     
\end{align}
Above we used the following relations for $\sigma^i$ being Pauli matrices for $i \in x,y,z$,
\begin{align}
    S_j^i &= \mathbb{1}_{1} \otimes \cdots \hspace{0.5em} \frac{\hbar}{2} \sigma^i_{j} \otimes \cdots \hspace{0.5em} {\mathbb{1}}_{N}, \\
    P^\dagger S_j^i S_{j+1}^i P &=  \prescript{}{1}{|-\rangle} \langle-|_1 \otimes \mathbb{1}_2 \cdots \hspace{0.2em} \frac{\hbar}{2} \sigma^i_{j} \otimes \frac{\hbar}{2} \sigma^i_{j+1} \cdots \hspace{0.2em} \prescript{}{N}{|-\rangle} \langle -|_N = S_j^i S_{j+1}^i P. 
\end{align}
For $j=1$ and $N$, the operators are therefore effectively replaced by their expectation values $\mathbf{\Tilde{S}}_1 = \frac{\hbar}{2}(1,0,0)$ and $\mathbf{\Tilde{S}}_N = \frac{\hbar}{2}(\cos{\phi}, \sin{\phi}, 0)  $. We arrive at the reduced effective Hamiltonian $H_{\text{eff}}$ of the main text by tracing out the edge spin states in the above term as those are fixed now, 
\begin{align}
    H_\text{eff} &= \Tr_{1,N}{(P^\dagger H(\phi) P)} = J \frac{\hbar}{2} (S_2^x  + \cos{\phi} S_{N-1}^x  + \sin{\phi} S_{N-1}^y ) +\sum_{j=2}^{N-2} \left[J(S_j^{x} S_{j+1}^{x} +S_{j}^{y} S_{j+1}^{y}) + \Delta S_{j}^{z} S_{j+1}^{z}\right] \\   
    H_\text{eff} &= J[\mathbf{\Tilde{S}}_1\cdot\mathbf{S}_2 + \mathbf{S}_{N-1}\cdot\mathbf{\Tilde{S}}_N(\phi)] +\sum_{j=2}^{N-2} \left[J(S_j^{x} S_{j+1}^{x} +S_{j}^{y} S_{j+1}^{y}) + \Delta S_{j}^{z} S_{j+1}^{z}\right], 
\end{align}
where the Hilbert space dimension is reduced by a factor of $4$ and $S_j^i+ = \mathbb{1}_{1} \otimes \cdots (\hbar/2) \sigma^i_{j-1} \otimes \cdots {\mathbb{1}}_{N-2}$ compared to the original chain of length $N$. 

\section{Adiabatic evolution with Kato's approach} \label{app:KatoApproach}

\noindent This section describes the numerical implementation of Kato's projection formalism for the adiabatic winding of phase $\phi$ of a particular state of the spin chain model mentioned in the main text. Following \cite{Kato1950OnTheAdiabaticTheoremOfQuantumMechanics}, we can write the full adiabatic evolution operator for a specific subspace as
\begin{align} 
    U_{\rm adia}(\phi) = \lim_{\delta \phi \rightarrow 0} \prod_{n=0}^{\phi/\delta \phi} V (n\cdot \delta \phi).
\end{align}
Here, $V(\phi)$ is the projection on the degenerate subspace corresponding to the eigenvalue $E_n(\phi)$ of $H(\phi)$. For our calculations, we implemented the repeated projection operators on the initial state, where one operator propagates by a small time step $\delta \phi = \pi/50$. For consecutive projections, we calculate the eigenspectrum and eigenvectors at the next successive phase $(\phi + \delta \phi)$ and calculate the overlap of the state (at $\phi$) with all eigenspaces at this next point. Then we select the subspace with maximum overlap as our adiabatically evolved subspace. For $\delta \phi \rightarrow 0$, this would be one, and overlap with other subspaces would be zero, but for finite $\delta \phi$ we fixed the minimum limit of $0.95$ for practical purposes. 

This projected subspace with a minimum overlap of $0.95$ with the last step's state is taken as the successor, and we iterate this step to reach the adiabatic evolved state at $\phi = \phi_f$ of initial state at $\phi = 0$. In case the maximum overlap is less than 0.95, we decrease the $\delta \phi$ dynamically at each step of iteration and increase it after the step again. The increase of $\delta \phi$ again helps to decrease the number of steps in the calculation without losing the projected subspace as we decrease $\delta \phi$ when the subspace does not have enough overlap.

\section{The adiabatic contribution to the spin current} \label{app:SpinPump}

\noindent In this section, we calculate the spin pumped in $z$-direction through our system by adiabatic rotation. To this end, we follow the ideas of Thouless \cite{Thouless1983PhysRevB.27.6083} and Fu and Kane \cite{FuKane2006PhysRevB.74.195312}. We see that spin current in the $z$-direction is conserved along the chain, i.e.,  
\begin{align}
    \langle \dot{S_j^z} \rangle &= \expval{i[H(t),S_j^z]} = 0,\\
    i[H(t),S_j^z] &= \underbrace{J \hbar (S_{j-1}^xS^y_{j}-S^y_{j-1}S^x_j)}_{C^z_{j-1,j}} - \underbrace{J \hbar (S_{j}^xS^y_{j+1}-S^y_{j}S^x_{j+1})}_{C^z_{j,j+1}},
\end{align}
where $C^z_{j,j+1}$ defines the spin current flowing from site $j$ to $j+1$ along the $z$-direction. For the last quantum spin, we have $\langle C^z_{N-1,N} \rangle = - \langle \partial_\phi H \rangle$, hence we have $\langle C^z_{N-2,N-1} \rangle = - \langle \partial_\phi H \rangle$. For the adiabatic evolution, the slowly changing ground state up to the first order excluding the instantaneous phase factor is given by \cite{Thouless1983PhysRevB.27.6083}
\begin{align}
     \label{eqn:ThoulessAdiabaticEvolution}
     |\psi_0(t)\rangle&\approx|\psi_0^{in}(t)\rangle+i \hbar \sum_{n\neq0}\frac{\langle\psi_n^{in}(t)|\dot{\psi}_0^{in}(t)\rangle}{\epsilon_n(t)-\epsilon_0(t)}|\psi_n^{in}(t)\rangle,
\end{align}
where the label ``in'' denotes the instantaneous eigenstate of $H(t)$, $\epsilon_n(t)$ is the energy of the $n^{th}$ eigenstate and ``$0$'' labels the ground state of $H(0)$. Here, the parallel transport gauge was used, in which the phase is chosen such that $\langle\psi_0^{in}(t)|\dot{\psi}_0^{in}(t)\rangle = 0$ \cite{Thouless1983PhysRevB.27.6083}. By \refEq{eqn:ThoulessAdiabaticEvolution}, we calculate the time-dependent $z$-spin current 
\begin{align}
    \langle\psi_0(t)|C^z_{N-2,N-1}|\psi_0(t)\rangle
    &=-\langle\psi_0^{in}(t)|\partial_\phi H(\phi)|\psi_0^{in}(t)\rangle\nonumber\\
    &+i \hbar \sum_{n\neq0}\frac{\langle\psi_0^{in}(t)|\partial_\phi H(\phi)|\psi_n^{in}(t)\rangle\langle\psi_n^{in}(t)|\dot{\psi}_0^{in}(t)\rangle}{\epsilon_n(t)-\epsilon_0(t)}\nonumber\\
    &-i\hbar \sum_{n\neq0}\frac{\langle\dot{\psi}_0^{in}(t)|\psi_n^{in}(t)\rangle\langle\psi_n^{in}(t)|\partial_\phi H(\phi)|\psi_0^{in}(t)\rangle}{\epsilon_n(t)-\epsilon_0(t)}\, \label{eq:FirstOrderCurrent}.
\end{align}
To solve this, we take the derivative of the energy eigenvalue equation 
\begin{align}
    \partial_\phi [H(\phi) |\psi_{q}^\text{in}(\phi) \rangle] =& \partial_\phi [\epsilon_q(\phi) | \psi_q^{in} (\phi) \rangle],  \nonumber \\
    [\partial_\phi H(\phi)] |\psi_{q}^\text{in}(\phi) \rangle + H(\phi) |\partial_\phi \psi_{q}^\text{in} (\phi) \rangle =& [\partial_\phi \epsilon_q(\phi)] | \psi_q^\text{in} (\phi) \rangle + \epsilon_q(\phi) |\partial_\phi \psi_{q}^\text{in} (\phi)\rangle.   
\end{align}
Taking the dot product with $\langle \psi_p^{in} (\phi) |, p \neq q$, we find 
\begin{align}
    \langle \psi_p^{in} (\phi) |[\partial_\phi H(\phi)] |\psi_{q}^{in} \rangle + \epsilon_p(\phi) \langle \psi_p^{in} (\phi) | \partial_\phi \psi_{q}^{in} (\phi) \rangle =& \hspace{0.5em} 0 + \epsilon_q(\phi) \langle \psi_p^{in} (\phi) |\partial_\phi \psi_{q}^{in} (\phi)\rangle, \nonumber \\ 
    \frac{\langle \psi_p^{in} (\phi) |[\partial_\phi H(\phi)] |\psi_{q}^{in} \rangle}{\epsilon_q(\phi) - \epsilon_p(\phi)} =& \langle \psi_p^{in} (\phi) |\partial_\phi \psi_{q}^{in} (\phi)\rangle. \label{eq:ThoulessCancel}
\end{align}
Substituting Eq.~(\ref{eq:ThoulessCancel}) into Eq.~(\ref{eq:FirstOrderCurrent}) and using $\sum_{n \neq 0} \ket{\psi_n^{in}(t)} \bra{{\psi_n^{in}(t)}} = \mathbb{1} - \ket{\psi_0^{in}(t)} \bra{\psi_0^{in}(t)}$ and for constant $\dot{\phi}$ we get
\begin{align}
    \langle\psi_0(t)|C^z_{N-2,N-1}|\psi_0(t)\rangle &= - \partial_\phi \epsilon_0(t) -i \hbar \sum_{n \neq 0} \langle \partial_\phi \psi_0^{in}(t) | \psi_n^{in}(t) \rangle \langle\psi_n^{in}(t)|\dot{\psi}_0^{in}(t)\rangle  + i \hbar \sum_{n \neq 0} \langle\dot{\psi}_0^{in}(t)|\psi_n^{in}(t)\rangle \langle \psi_n^{in}(t) | \partial_\phi \psi_0^{in}(t) \rangle, \nonumber \\
    &= - \partial_\phi \epsilon_0(t) -i \hbar \dot{\phi}\underbrace{[ \langle \partial_\phi \psi_0^{in}(t)|\partial_\phi \psi_0^{in}(t) \rangle - \langle \partial_\phi \psi_0^{in}(t) |\partial_\phi \psi_0^{in}(t) \rangle]}_{=0}   \nonumber \\ 
    & +i \hbar \dot{\phi}\underbrace{(\langle \partial_\phi \psi_0^{in}(t)| \psi_0^{in}(t) \rangle \langle \psi_0^{in}(t)| \partial_\phi \psi_0^{in}(t) \rangle - \langle \partial_\phi \psi_0^{in}(t)| \psi_0^{in}(t) \rangle \langle \psi_0^{in}(t)| \partial_\phi \psi_0^{in}(t) \rangle )}_{=0}.  \label{eqZero}
\end{align}
Here, $-\partial_\phi \epsilon_0(t)$ corresponds to the instantaneous spin current in the eigenstate, while additional terms correspond to a topological adiabatic contribution of the pumped $z$-spin current which vanishes at all phases.
  
\section{Overlap of the adiabatically wound-up states with phantom states} \label{app:Overlaps}

\noindent The phantom helix states for the $XXZ$ Heisenberg model of the main text are \cite{Kuehn2023}
\begin{align}
    |PH\rangle^{\pm}_{1} =& \frac{1}{\sqrt{2^N}} \bigotimes_{j=2}^{N-1}(|\uparrow \rangle_j - e^{\pm i (j-2)\gamma} |\downarrow \rangle_j ), \label{eq:PhantomStates1} \\
    |PH\rangle^{\pm}_{2} =& \frac{1}{\sqrt{2^N}} \bigotimes_{j=2}^{N-1}(|\uparrow \rangle_j + e^{\pm i j\gamma} |\downarrow \rangle_j ), \label{eq:PhantomStates2}
\end{align}
which are subject to the phantom condition for the boundary phase
\begin{align}
    \phi_{ph} = \pm (N-M)\pi/3 + \pi \delta_{3,M}  \text{ mod } 2\pi. \label{eq:PhantomBoundary}
\end{align} 
The $\pm$ in \refEq{eq:PhantomStates1} and (\ref{eq:PhantomStates2}) corresponds to the boundary conditions in \refEq{eq:PhantomBoundary}, $M$ being $5$ or $3$ for $|PH\rangle^{\pm}_{1}$ and $M$ being $3$ or $1$ for $|PH\rangle^{\pm}_{2}$. The $|\uparrow \rangle_j$ and $|\downarrow \rangle_j$ states are up and down state in $z$-direction at site $j$.
\refTab{tab:Overlap} enlists the overlaps of the phantom state with the adiabatically evolved state when the phantom boundary conditions are fulfilled. The overlap is $O = |\prescript{\pm}{1,2}{\langle} PH | \hspace{0.4em} U_{\rm adia}(\phi) | GS \rangle |^2$,  where $\phi$ and $|GS \rangle$ is the corresponding phantom boundary phase and ground state, respectively.  It is seen that the overlap decreases with the increase in chain length although the subspace for both states is always the same. 

\begin{table*}[ht]
    \centering
    \begin{tabular}{||m{5em}|m{6em}|m{6em}|m{6em}||m{5em}|m{6em}|m{6em}|m{6em}||}
        \hline
         \textbf{Chain Length $N$} & \textbf{Phantom State $|PH\rangle$} & \textbf{Boundary Phase $\phi$} & \textbf{Overlap $O$} & \textbf{Chain Length $N$} & \textbf{Phantom State $|PH\rangle$} & \textbf{Boundary Phase $\phi$} & \textbf{Overlap $O$} \\ \hline \hline   
         
         \multirow{8}{5em}{\textbf{5}} & \multirow{2}{6em}{$|PH\rangle^{+}_{1}$} &  $5\pi /3$  & 0.82468 & \multirow{8}{5em}{\textbf{6}} & \multirow{2}{6em}{$|PH\rangle^{+}_{1}$} &  $\pi/3 + 2\pi$  & 0.95907 \\ \cline{3-4} \cline{7-8}
          & & $2\pi$& 0.98483 & & & $2\pi$& 0.75917 \\ \cline{2-4} \cline{6-8}
          & \multirow{2}{6em}{$|PH\rangle^{+}_{2}$} & $4\pi /3$
          & 0.95913 & & \multirow{2}{6em}{$|PH\rangle^{+}_{2}$} & $5\pi /3$ & 0.92944 \\ \cline{3-4} \cline{7-8}
          & & $5\pi /3$ & 0.82465 & & & $2\pi$& 0.80013\\ \cline{2-4} \cline{6-8}
          & \multirow{2}{6em}{$|PH\rangle^{-}_{1}$} & $6\pi$ & 0.98307 & & \multirow{2}{6em}{$|PH\rangle^{-}_{1}$} & $5\pi /3 + 2\pi$ & 0.95907 \\ \cline{3-4} \cline{7-8}
          & & $\pi/3 + 6\pi$ & 0.82468 & & & $4\pi$ & 0.80029\\ \cline{2-4} \cline{6-8}
          & \multirow{2}{6em}{$|PH\rangle^{-}_{2}$} & $\pi/3 + 6\pi$ & 0.82465 & & \multirow{2}{6em}{$|PH\rangle^{-}_{2}$} & $4\pi$ & 0.75900 \\ \cline{3-4} \cline{7-8}
          & & $2\pi /3 +6\pi$ & 0.95913 & & & $\pi/3 + 4\pi$ & 0.92944\\ \hline \hline

         \multirow{8}{5em}{\textbf{7}} & \multirow{2}{6em}{$|PH\rangle^{+}_{1}$} &  $\pi/3 +2\pi$  & 0.73919  & \multirow{8}{5em}{\textbf{8}} & \multirow{2}{6em}{$|PH\rangle^{+}_{1}$} &  $\pi +2\pi$  & 0.90263 \\ \cline{3-4} \cline{7-8}
          & &$2\pi /3 + 2\pi$ & 0.92938  & & &$2\pi /3 +2\pi$ & 0.70163 \\ \cline{2-4} \cline{6-8}
          & \multirow{2}{6em}{$|PH\rangle^{+}_{2}$} & $2\pi$ & $0.89156$ & & \multirow{2}{6em}{$|PH\rangle^{+}_{2}$} & $\pi /3 +2\pi$ & 0.86400 \\ \cline{3-4} \cline{7-8}
          & & $\pi/3 + 2\pi$ & 0.73915  & & & $2\pi /3 +2\pi$& 0.70159 \\ \cline{2-4} \cline{6-8}
          & \multirow{2}{6em}{$|PH\rangle^{-}_{1}$} & $4\pi /3 + 8\pi$ & 0.92938 & & \multirow{2}{6em}{$|PH\rangle^{-}_{1}$} & $\pi +4\pi$ & 0.89175 \\ \cline{3-4} \cline{7-8} 
          & & $5\pi /3 + 8\pi$ & 0.73919  & & &$4\pi /3 +4\pi$ & 0.70163 \\ \cline{2-4} \cline{6-8}
          & \multirow{2}{6em}{$|PH\rangle^{-}_{2}$} & $5\pi /3 + 8\pi$ & 0.73915  & & \multirow{2}{6em}{$|PH\rangle^{-}_{2}$} & $4\pi /3 +4\pi$ & 0.70175 \\ \cline{3-4} \cline{7-8} 
          & & $10\pi$ & 0.90289  & & & $5\pi / 3 +4\pi$ & 0.86400 \\ \hline \hline

         \multirow{8}{5em}{\textbf{9}} & \multirow{2}{6em}{$|PH\rangle^{+}_{1}$} &  $\pi +2\pi$  & 0.64146  & \multirow{8}{5em}{\textbf{10}} & \multirow{2}{6em}{$|PH\rangle^{+}_{1}$} &  $4\pi /3 +2\pi$  & 0.63394  \\ \cline{3-4} \cline{7-8}
          & & $4\pi /3 +2\pi$ & 0.86394  & & & $5\pi /3 +2\pi$ & 0.83039  \\ \cline{2-4} \cline{6-8}
          & \multirow{2}{6em}{$|PH\rangle^{+}_{2}$} & $2\pi/3 +2\pi$ & 0.83045  & & \multirow{2}{6em}{$|PH\rangle^{+}_{2}$} & $\pi  +2\pi$ & 0.78771 \\ \cline{3-4} \cline{7-8}
          & & $\pi +2\pi$ & 0.69150 & & & $4\pi /3 +2\pi$ & 0.63389\\ \cline{2-4} \cline{6-8}
          & \multirow{2}{6em}{$|PH\rangle^{-}_{1}$} & $2\pi /3 +12\pi$ & 0.86394  & & \multirow{2}{6em}{$|PH\rangle^{-}_{1}$} & $\pi /3 +6\pi$ & 0.83039 \\ \cline{3-4} \cline{7-8}
          & & $\pi +12\pi$ & 0.69172 & & & $2\pi /3 +6\pi$ & 0.63394 \\ \cline{2-4} \cline{6-8}
          & \multirow{2}{6em}{$|PH\rangle^{-}_{2}$} & $\pi +12\pi$ & 0.64172 & & \multirow{2}{6em}{$|PH\rangle^{-}_{2}$} & $2\pi /3 +6\pi$ & 0.63389 \\ \cline{3-4} \cline{7-8}
          & & $4\pi /3 +12\pi$ & 0.83045 & & & $ \pi  +6\pi$ & 0.80640 \\ \hline \hline
          
    \end{tabular}
    \caption{List of phantom state overlaps with the adiabatically evolved state for the respective chain length.}
    \label{tab:Overlap}
\end{table*}

\section{Fits of the different energy scalings of the adiabatic curve with the chain length}  

\label{app:Fits}
\begin{figure*}[b]
    \vspace{-1em}
    \centering
    \setlength{\unitlength}{0.1\textwidth}
    \begin{picture}(10,3.5)
        \put(0,0){\includegraphics[width=0.98\textwidth]{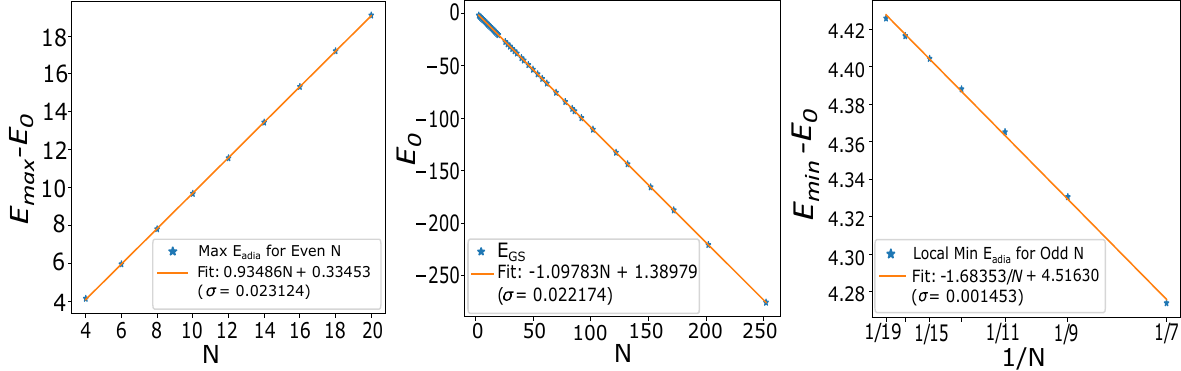}}
        \put(0.57,3.1){\textbf{(a)}}
        \put(3.85,3.1){\textbf{(b)}}
        \put(7.2,3.1){\textbf{(c)}}
    \end{picture}
    \caption{Scaling of the energy vs.~the system size $N$ including the fit shown in the respective inset. Panel (a) shows the linear fit of the increase in the maximum pump of adiabatic energy w.r.t. the ground state energy, and the decrease of ground state energy is shown in panel (b). Panel (c) shows the $1/N$ fitting of center minima for the odd chains adiabatic curve.} 
    \label{fig:Scaling}
    \vspace{-1em}
\end{figure*}

\noindent In this section, we give details for the linear fit of the maximum adiabatic pumped energy and of the ground state energy w.r.t.~the chain length, shown in Fig.~\ref{fig:Scaling}(a) and (b). The least-square fit shows the adiabatic energy pump scales with $E_{max}-E_0 \approx 0.93486N+0.33453$ and the ground state energy decreases linearly with $E_{0}\approx -1.09783N + 1.38979$. We calculate the ground state energy using the Density Matrix Renormalization Group (DMRG) method. For odd chain lengths, we also performed the fitting for the local minima situated at the center of the adiabatic curve to get the curve of $E_{min} - E_0 \approx -1.68353 / N + 4.51630$ shown in Fig.~\ref{fig:Scaling}(c).   

\section{Permutation cycles of adiabatic twisting} \label{app:PermutaionCycles}
\noindent After adiabatically rotating one boundary spin once, we reach the same Hamiltonian $H(0) = H(2\pi)$. We calculate the adiabatic evolution for all states from $\phi=0$ to $\phi=2\pi$ and define the cyclic permutations describing how the eigenstates are transformed into each other after one rotation. From each cycle's length, we infer each state's periodicity and find the super-winding states. \refTab{tab:Cycles} shows the lists of cycles numbering the eigenstates representing the eigenstates that are reached after one consecutive rotation of the edge spin for $\Delta = -|J|/2$. It also shows each cycle's length, corresponding to the periodicity of these states in the adiabatic winding process. From that, we can see the spin chain with length $N=8$ has the highest periodic state of periodicity $6$, and of length $N=9$ has the highest periodic state of periodicity $12$. Data for more chain lengths is available on request. 
\begin{table*}[ht]
    \centering
    
    \text{\textbf{(a)} Chain Length = 8}
    \begin{tabular}{|c|c|c|c|c|c|c|c|c|}
        \hline
        \textbf{4} [1, 8, 38, 5]&
        \textbf{4} [2, 12, 29, 9]&
        \textbf{4} [3, 11, 21, 10]&
        \textbf{6} [4, 18, 24, 19, 25, 15]&
        \textbf{4} [6, 7, 17, 16]&
        \textbf{3} [13, 26, 23]&
        \textbf{1} [14]&
        \textbf{3} [20, 37, 34]\\ \hline
        \textbf{1} [22]&
        \textbf{3} [27, 46, 43]&
        \textbf{3} [28, 36, 35]&
        \textbf{3} [30, 54, 50]&
        \textbf{1} [31]&
        \textbf{3} [32, 45, 44]&
        \textbf{2} [33, 39]& 
        \textbf{3} [40, 53, 51]\\ \hline
        \textbf{1} [41]&
        \textbf{2} [42, 49]&
        \textbf{1} [47]&
        \textbf{2} [48, 56]&
        \textbf{2} [52, 58]&
        \textbf{2} [55, 61]&
        \textbf{1} [57]&
        \textbf{2} [59, 63]\\ \hline
        \textbf{1} [60]&
        \textbf{1} [62]&
        \textbf{1} [64]&
        &
        &
        &
        &
        \\ \hline  
    \end{tabular}
    
    \vspace{1em}
    \text{\textbf{(b)} Chain Length = 9}
    \begin{tabular}{|c|c|c|c|c|c|c|c|c|}
        \hline
        \textbf{8} [1,8,64,44,23,47,60,5]&
        \textbf{8} [2,13,53,29,10,33,52,9]&
        \textbf{8} [3,15,40,35,20,36,38,14]&
        \textbf{4} [4,21,50,17]\\ \hline
        \textbf{12} [6,12,28,30,26,32,27,11,7,24,37,22]&
        \textbf{4} [16,19,25,18]&
        \textbf{6} [31,49,57,42,58,48]&
        \textbf{6} [34,66,94,72,96,61]\\ \hline
        \textbf{4} [39,46,54,45]&
        \textbf{6} [41,80,105,83,107,76]&
        \textbf{6} [43,68,73,59,74,67]&
        \textbf{6} [51,88,110,97,111,84]\\ \hline
        \textbf{4} [55,63,71,62]&
        \textbf{6} [56,82,92,70,95,81]&
        \textbf{6}[65,93,101,79,102,90]&
        \textbf{4} [69,78,89,77]\\ \hline
        \textbf{4} [75,87,98,85]&
        \textbf{2} [86,91]&
        \textbf{2} [99,103]&
        \textbf{4} [100,116,117,115]\\ \hline
        \textbf{2} [104,109]&
        \textbf{4} [106,122,120,121]&
        \textbf{2} [108,114]&
        \textbf{4} [112,126,124,125]\\ \hline
        \textbf{2} [113,118]&
        \textbf{2} [119,123]&
        \textbf{2} [127,128]&
        \\ \hline  
    \end{tabular}
     
    \caption{List of permutation cycles after $2\pi$ adiabatic rotation at $\Delta = -|J|/2$ with the cycle length corresponding to those states' periodicity. [$ n_1, n_2, \cdots$] denotes the cycle, listing the $n^{th}$ eigenstates, and the periodicity is written in bold for the chain length of (a) $N=8$ and (b) $N=9$.  The eigenstates [4,18,24,19,25,15] with $Z_6$ periodicity form the largest cycle for $N=8$ and eigenstates [6,12,28,30,26,32,27,11,7,24,37,22] with $Z_{12}$ periodicity form largest cycle for $N=9$.}
    \label{tab:Cycles}
    \vspace{-2em}
\end{table*}

\end{document}